\documentclass[superscriptaddress,aps,twocolumn]{revtex4}
\usepackage{braket}
\usepackage{graphicx}
\usepackage{amsmath}
\usepackage{amsfonts}
\usepackage{chngcntr}
\usepackage{float}
\usepackage{comment}
\usepackage{ulem}
\usepackage[dvipsnames]{xcolor}
\usepackage{soul}
\usepackage{hyperref}
\usepackage{comment}
\usepackage{array}

\DeclareMathOperator{\Tr}{Tr}

\begin{document}
\title{Existence and characterization of edge states\\ in an acoustic trimer Su-Schrieffer-Heeger model}

\author{I.~Ioannou Sougleridis}  
\affiliation{Laboratoire d’Acoustique de l’Universit\'{e} du Mans (LAUM), UMR 6613, Institut d'Acoustique - Graduate School (IA-GS), CNRS,  Le Mans Universit\'{e}, France}
\affiliation{Department of Physics, National and Kapodistrian University of Athens, University Campus, GR-157 84 Athens, Greece}
\author{A.~Anastasiadis}  
\affiliation{Laboratoire d’Acoustique de l’Universit\'{e} du Mans (LAUM), UMR 6613, Institut d'Acoustique - Graduate School (IA-GS), CNRS,  Le Mans Universit\'{e}, France}
\author{O.~Richoux}  
\affiliation{Laboratoire d’Acoustique de l’Universit\'{e} du Mans (LAUM), UMR 6613, Institut d'Acoustique - Graduate School (IA-GS), CNRS,  Le Mans Universit\'{e}, France}
\author{V.~Achilleos}  
\affiliation{Laboratoire d’Acoustique de l’Universit\'{e} du Mans (LAUM), UMR 6613, Institut d'Acoustique - Graduate School (IA-GS), CNRS,  Le Mans Universit\'{e}, France}
\author{G.~Theocharis}  
\affiliation{Laboratoire d’Acoustique de l’Universit\'{e} du Mans (LAUM), UMR 6613, Institut d'Acoustique - Graduate School (IA-GS), CNRS,  Le Mans Universit\'{e}, France}
\author{V.~Pagneux}  
\affiliation{Laboratoire d’Acoustique de l’Universit\'{e} du Mans (LAUM), UMR 6613, Institut d'Acoustique - Graduate School (IA-GS), CNRS,  Le Mans Universit\'{e}, France}
\author{F.K.~Diakonos}  
\affiliation{Department of Physics, National and Kapodistrian University of Athens, University Campus, GR-157 84 Athens, Greece}

\begin{abstract}
We report on a direct mapping of acoustic slender waveguides to the one dimensional trimer Su-Schrieffer-Heeger model, with neither chiral nor mirror symmetry. Importantly, we can choose to perform this mapping for either the acoustic velocity or pressure. 
We demonstrate that, for finite systems, this choice is necessarily linked to the boundary conditions. 
It allows for the unveiling of the edge states of the acoustic system through an edge state phase diagram. An experimental realization of our setup in the audible regime corroborates our theoretical predictions. 


\end{abstract}

\maketitle





\section{Introduction}
\normalem

Topology has a high impact on the design of a new class of wave systems in many different areas, such as photonics \cite{photonics,photonics2}, phononics \cite{mechanical,phononic} and acoustics \cite{ma,review_acoustics}. One of the fundamental properties of topological systems is that they exhibit robust localized states on their boundaries or interfaces. In this context, the most renowned one-dimensional (1D) model 
is the Su-Schrieffer-Heeger (SSH) dimer model \cite{su,asboth}, which appears as a prototype for topological insulators due to its richness and simplicity, featuring
robust localized edge modes at zero energy.

Among the many extensions of the SSH model that exist both in one  and higher-dimensions \cite{franke2018,groning2018,rizzo2018,coupled_ssh,wakabayashi,photonics_2D,zheng2019,xiao2017,liao2022,kaleidoscope,huda2020tuneable}, the SSH trimer (SSH3) model has been of particular interest in the recent years \cite{huda2020tuneable,alvarez2019,coutinho,  super_SSH,floquet_trimer,extended_ssh,anastasiadis2022bulk,kaleidoscope,shi2023,decorated_trimer,trimer_soliton,rotation_trimer}. The peculiarity of the SSH3 is that, in contrast to SSH, it is missing chiral symmetry and occasionally also mirror symmetry.
   \,Therefore, in general, the topological invariants of the SSH, namely, the winding number and the Zak phase, are not quantized for SSH3 and they consequently fail to establish a coherent bulk-boundary correspondence (BBC) as in the case of the SSH model. Despite the lack of those symmetries, the SSH3 exhibits robust localized modes (which no longer manifest at zero energy). Nevertheless, an alternative 1D bulk quantity called the normalized sublattice Zak phase was recently proposed to establish BBC \cite{anastasiadis2022bulk}. 

In this article, we introduce a periodic configuration of different slender acoustic waveguides of the same length, similar to  \cite{zheng2019,coutant_corner,coutant_robustness,coutant2021acoustic,coutant2022}, 
as an acoustic analogue of the SSH3 lattice using either acoustic pressure or velocity. 
 The associated couplings
are given as analytical expressions of the waveguide cross-sections. Two finite configurations are considered: i)\,firstly, an open structure for which the mapping is established by using the pressure as the dependent variable and ii)\,secondly, a closed structure for which the mapping is established by using the velocity. As recently shown in \cite{allein2022strain} for systems described by \emph{mass-spring} models, the appropriate choice of dependent variable, depending on the boundary conditions, is crucial to reveal the underlying topological characteristics. Here, we demonstrate that the appropriate choice of dependent variables is also relevant to predict the number of edge states in continuous \emph{acoustic} systems. In particular, we translate the edge state phase diagram of the SSH3 \cite{anastasiadis2022bulk} to the acoustic configurations for both open and closed boundaries. We confirm our theoretical predictions through appropriate experimental setups, featuring different numbers of edge states.

\section{The trimer Su-Schrieffer-Heeger model (SSH3)}\label{sec1}





The SSH3 model is a periodic lattice whose unit cell consists of three sites connected by three alternating couplings $u,v$ (intra-cell)
and $w$ (inter-cell) as is shown in Fig.\,\ref{fig1}\,\textbf{(a)}. Due to the periodicity of the model, Bloch theorem can be applied and the Bloch Hamiltonian (alternatively: bulk Hamiltonian) of this system takes the form:


\begin{figure*}
\includegraphics[width=\textwidth]{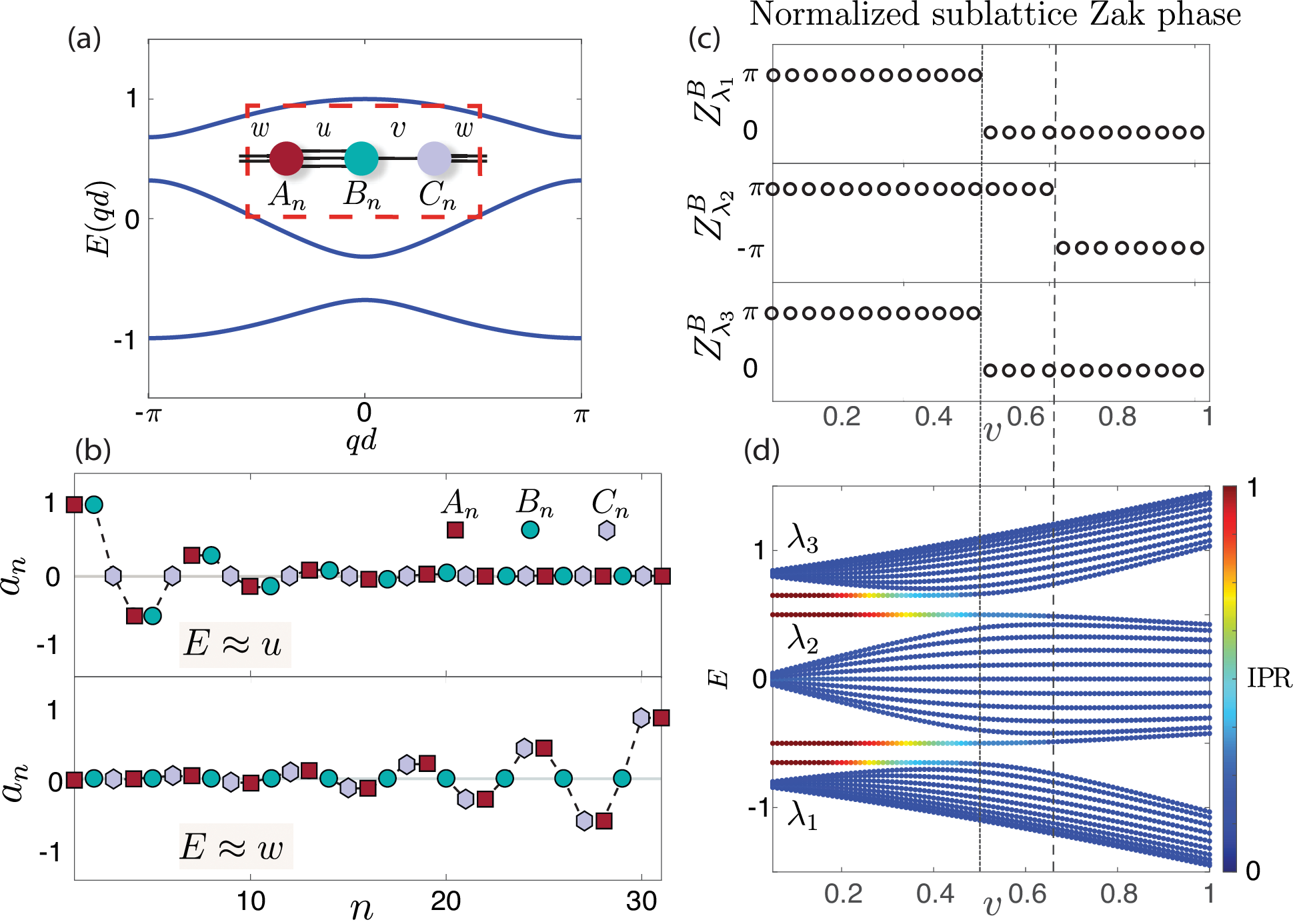}
 \caption{ \textbf{The SSH3 model} -  \textbf{a)} Unit cell of SSH3 and its band structure. Three different couplings $u,v$ and $w$ connect the lattice sites. In all of the sub-figures the coupling values are $u=0.4804,\, w=0.6591$, and if not mentioned otherwise $v=0.3430$. \textbf{b)} The profile of two edge states appearing in the first band gap exhibited by a finite SSH3 consisting of 31 sites. 
 \textbf{c)} Calculation of $Z_{\lambda_i}^B$ for each band ($\lambda_1,\,\lambda_2$ and $\lambda_3$) as we change the value of the coupling $v$ while keeping constant $u$ and $w$. \textbf{d)} The spectrum of a finite SSH3 consisting of 31 sites (non-integer number of unit cells) as we change $v$, while keeping constant other couplings. The colorbar illustrates the inverse participation ratio of the eigenmodes, for each value of $v$. Each jump of $\pi$ in normalized Zak phase (dotted lines) in \textbf{c)} corresponds to an emergence of edge state in \textbf{d)}.
}\label{fig1}
\end{figure*}

\begin{align}\label{ss3_ham}
    H(q) = \begin{pmatrix}
    0 & u & we^{-iqd}\\
    u & 0 & v\\
    we^{iqd} & v &0
    \end{pmatrix}
\end{align}
where $q\in [-\pi/d, \pi/d)$ is the Bloch wavenumber and $d$ is the length of the unit cell. The system exhibits a band structure similar to the one shown in Fig.\,\ref{fig1}\,\textbf{(a)} where one can observe that the bands are symmetric with respect to the point $(\pi/2,0)$ on the $(E,q)$ plane. 
%

Furthermore, a BBC was recently established for this model \cite{anastasiadis2022bulk}. BBC correlates the number of edge states exhibited in a finite system (broken translation invariance) to the values of a bulk quantity defined on the infinite (bulk) system. A well-defined bulk quantity for this model is the normalized sublattice Zak phase

\begin{align}\label{sublattice_zak}
Z_{\lambda_j}^{r} := \frac{i}{2}\oint dq \braket{\tilde{a}^r_{\lambda_j}(q)|{\partial_q\tilde{a}^r_{\lambda_j}(q) }} ,
\end{align}
where $\lambda_i$ denotes the band which is probed and $r = A,B,C$ the sublattice. In Eq.\,\eqref{sublattice_zak} we used $\ket{\tilde{a}_{\lambda_j}^r(q)}~=~\displaystyle{\frac{P_{r}\ket{a_{\lambda_j}(q)}}{\sqrt{\bra{a_{\lambda_j}(q)}P_{r}\ket{a_{\lambda_{j}}(q)}}}}$ where $\ket{a_{\lambda_j}(q)}$ is the cell-periodic part of the Bloch wavefunctions and $P_{r}~=~\ket{r}\bra{r}$ is the projector to the $r$ sublattice. 
In this framework, the emergence of edge states for different lengths of the chain (3N sites, 3N+1 sites etc) can be predicted by the normalized sublattice Zak phase. Furthermore, this system can host 0,\,1 or 2 pairs of edge states for different values of the couplings. 
The number of edge states for the same coupling values, is depending to which sublattice belong the first and last sites of the chain (see Appendix \ref{appendix0}). In this paper we will examine theoretically and experimentally a chain with $3N+1$ sites, for which the first site belongs to sublattice $A$ and consequently the last site belongs to $A$ as well. 
The utility of $Z_{\lambda_j}^{r}$ is that one can predict the phase diagram for all of these cases by choosing the appropriate sublattice for its evaluation. For that case, we will use the normalized sublattice Zak phase $Z_{\lambda_j}^{B}$, since it has been previously shown to be a relevant 1D bulk quantity (see Eq. (21) in \cite{anastasiadis2022bulk}).
We find that the energies of the edge modes are approximately equal to $\pm u$ and $\pm w$, (for more details see Appendix \ref{appendix0}).
 Fig.\,\ref{fig1}\,\textbf{(b)} illustrates the aforementioned edge modes for a chain consisting of 31 sites (10 unit cells + 1 site). Notice that the dependent variable on sublattice $C$ for $E\approx u$ ($B$ for $E\approx w$ ) is approximately zero, as discussed in Appendix \ref{appendix0}. In Fig.\,\ref{fig1}\,\textbf{(c)} we plot the values of $Z_{\lambda_j}^{B}$ for each band of an SSH3 with coupling values $u=0.4804$ and $w=0.6591$, as we adiabatically change the value of $v$ from $0$ to $1$. We observe that $Z_{\lambda_1}^{B} $ and $Z_{\lambda_{3}}^{B} $ exhibit a "jump" from $\pi$ to $0$ around the point $v=u$, while $Z_{\lambda_2}^{B} $ exhibits a similar jump from $\pi$ to $-\pi$ around the point $v=w$. By comparing the values of this quantity to the continuation of the spectrum of a finite SSH3, 
 %
 illustrated in Fig \ref{fig1}\textbf{(d)}, we observe that edge states emerge from each band when a "jump" of the bulk quantity occurs. The number of edge states leaving each band is $N_{\lambda_i,edge} = |\Delta Z_{\lambda_i}/\pi|$ \cite{anastasiadis2022bulk}. The colorbar presents the inverse participation ratio (IPR)\footnote{The formula used to calculate inverse participation ratio is $IPR= \sum_{j=1}^N|a_j|^4/(\sum_{j=1}^N |a_j|^2)^2$, where $a_j$ are the normalized eigenmodes of the chain.}\par 


\section{Acoustic Analog of SSH3}
\subsection{From a 2D acoustic periodic waveguide to the SSH3 model}
\begin{figure*}[ht]
\begin{center}
\includegraphics[width=0.75\textwidth]{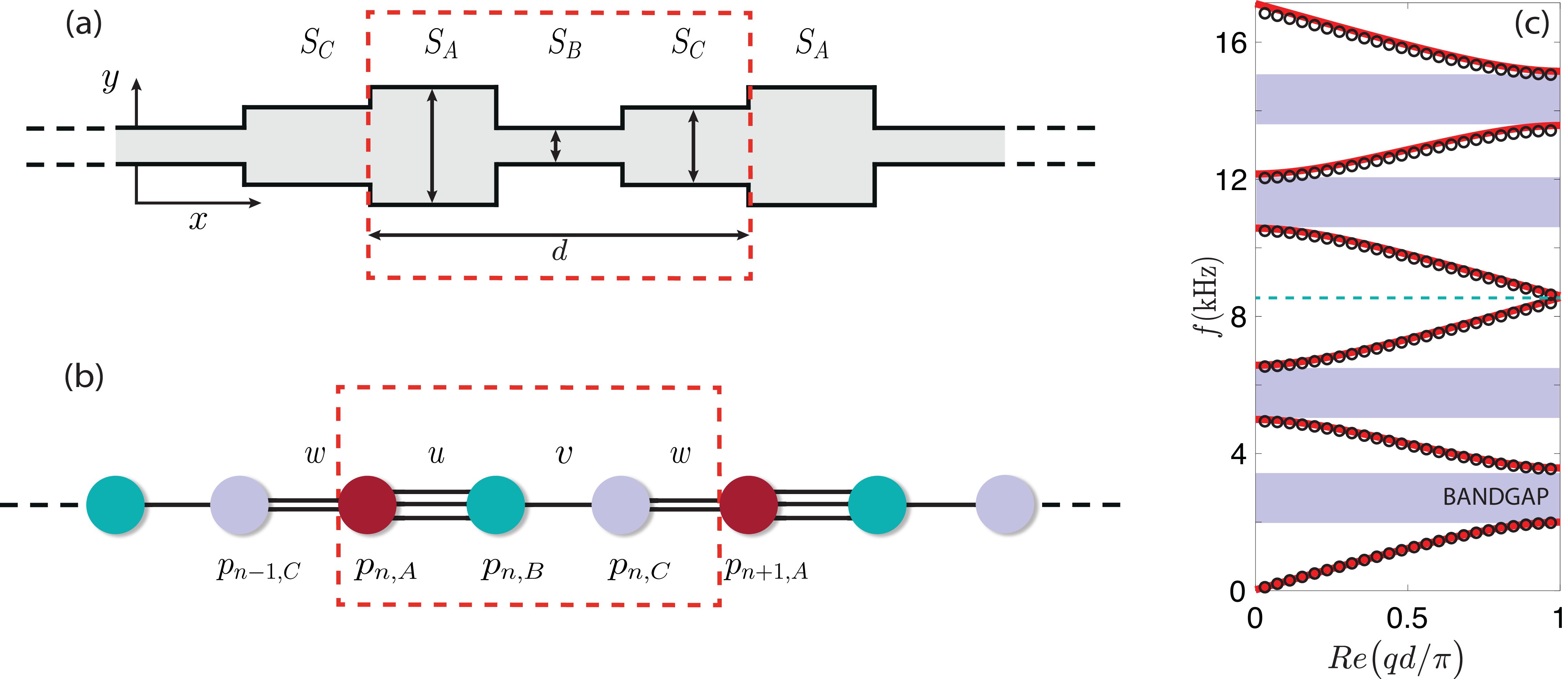}
 \caption{\textbf{The acoustic SSH3 analog} - \textbf{a)} The acoustic analog of the SSH3, a periodic waveguide with three alternating cross sections $S_A, \, S_B$ and $S_C$. \textbf{b)} Mapping to SSH3. The sites of SSH3 correspond to the "jump" points where the cross-sections change \textbf{c)} Dispersion relation of the waveguide obtained by
the TMM (circles) vs FEM (red line). Stars indicate the frequencies/pseudo-energies where the in-gap edge states are expected.   \label{fig2}}
\end{center}
\end{figure*}

We proceed now with the theoretical implementation of an acoustic waveguide system that can be mapped to the SSH3 model. First, we demonstrate that a 2D acoustic waveguide composed by segments of equal length $L$ and three different cross sections $S_A$, $S_B$ and $S_C$, arranged periodically with period $d=3L$, as depicted in Fig. \ref{fig2}\textbf{(a)} is a good platform to test the theoretical predictions for this model. We begin by showcasing how low frequency sound wave propagation can be modelled by a discrete set of equations through the transfer matrix method (TMM).

Our starting point is the consideration of an ideal fluid neglecting nonlinearity, viscosity and other dissipative
terms. In this regime, the acoustic pressure field $p(x,y)$ inside the waveguide of Fig. \ref{fig2}\textbf{(a)} is governed by the two-dimensional (2D) Helmholtz equation:
\begin{equation}
\frac{\partial^{2} p}{\partial x^{2}}+\frac{\partial^{2} p}{\partial y^{2}}+k^{2} p=0,
\label{helmholtz}
\end{equation}
 with Neumann conditions corresponding to zero normal velocity at the rigid walls $\partial_n p=0$. 
Here $k= \omega /c_0$ with $\omega$ the
angular frequency and $c_0$ the speed of sound.
Moreover, we are interested in sufficiently low frequencies such that only the fundamental acoustic mode of the waveguide is propagating (monomodal approximation). In this regime, the acoustic wave propagation is simply described by the one-dimensional (1D) Helmholtz equation \cite{coutant2021acoustic,Morse}

\begin{align}\label{Helmholtz_1d}
    \frac{d^2 p}{d x^2} + k^2 p = 0.
\end{align}

For the mapping, we follow exactly the same technique as in \cite{coutant2021acoustic}. Essentially, the sublattices of the SSH3 are mapped to the \emph{equidistant} points $x_{n,A}=nL$, $x_{n,B}=(n+1)L$ and $x_{n,C}=(n+2)L$ where the cross sections of the waveguide change as depicted in Fig.\,\ref{fig2}\,\textbf{(a-b)}. We denote the pressure amplitude at these points $p_{n,j\pm} \equiv p(x_{n,j\pm})$ with $j=A,B,C$. Next, we employ the jump conditions \eqref{jump_cond_1} enforcing the continuity of pressure and flux velocity at the changes of cross section $x_{n,j}$
\begin{align}\label{jump_cond_1}
    p_{n,j+}=p_{n,j-}, \quad  \phi_{n,j+}=\phi_{n,j-},
\end{align}
where $\phi_{n,j} := S_j \frac{d p_{n,j}}{dx}$ is the flux velocity at the change of cross section. By employing \eqref{jump_cond_1} and the piece-wise solution of the 1D Helmholtz Eq. \eqref{Helmholtz_1d}, some algebra leads to the system of equations for the renormalized pressure $\tilde{p}_{n,j}$ (see Appendix \ref{appendixa} for more details),



%

\begin{align}
&w \tilde{p}_{n-1,C}+u \tilde{p}_{n,B}=\tilde{p}_{n,A} E(k), \label{normalized11} \\
&u\tilde{p}_{n,A}+v \tilde{p}_{n,C}=\tilde{p}_{n,B}E(k), \label{normalized22}\\ 
&v \tilde{p}_{n,B}+w\tilde{p}_{n+1,A}=\tilde{p}_{n,C} E(k), \label{normalized33}  \\\nonumber
\end{align}
where 
\begin{equation}
   E(k) = \cos{kL}
   \label{pseudoEnergy}
\end{equation}
is a "pseudo-energy" and the couplings $u,v$ and $w$ are given as functions of the cross-sections $S_A$, $S_B$ and $S_C$
\begin{align}
&u=\frac{S_A}{\sqrt{\left(S_{C}+S_{A}\right)\left(S_{A}+S_{B}\right)}},\label{open couplings1}\\
&v=\frac{S_B}{\sqrt{\left(S_{B}+S_{C}\right)\left(S_{A}+S_{B}\right)}},\label{open couplings2}\\
&w =\frac{S_C}{\sqrt{\left(S_{C}+S_{A}\right)\left(S_{C}+S_{B}\right)}}\label{open couplings3}.
\end{align}
We note that the normalization is performed to recover the exact form of the system of discrete equations of the SSH3 (see Appendix \ref{appendixa}).
Furthermore, by using Bloch's theorem, 
$\tilde{p}_{n,A} = \tilde{p}_{A}e^{iqnd}$, $\tilde{p}_{n,B} = \tilde{p}_{B}e^{iqnd}$, $\tilde{p}_{n,C}= \tilde{p}_{C}e^{iqnd}$, one can recover the following eigenvalue problem

\begin{align}\label{acoustic_ssh3}
   \begin{pmatrix}
   0 & u & we^{-iqd}\\
   u & 0 & v \\
   we^{iqd} & v & 0 
   \end{pmatrix} \begin{pmatrix}
   \tilde{p}_{A}\\
   \tilde{p}_{B}\\
   \tilde{p}_{C}
   \end{pmatrix} = E(k) \begin{pmatrix}
   \tilde{p}_{A}\\
   \tilde{p}_{B}\\
   \tilde{p}_{C}
   \end{pmatrix}.
\end{align}
 By comparing \eqref{acoustic_ssh3} with \eqref{ss3_ham} we see that the matrix in \eqref{acoustic_ssh3} has the same form as the SSH3 Hamiltonian of equation \eqref{ss3_ham}. 
 
The mapping of pressure at the changes of cross section is ideal for the implementation of open ends (boundary conditions) as discussed in \cite{coutant2021acoustic}. However, bearing in mind an experimental investigation, in which closed ends are preferred to avoid radiation losses, we demonstrate that a similar mapping can be established for the flux velocity at the changes of cross section. As we did for the pressure mapping we use the piece-wise solution of the 1D Helmholtz Eq. for the flux velocity and apply the continuity of pressure and flux velocity at the changes of cross-section \eqref{jump_cond_1}, to obtain the following system of equations

\begin{align}
&w^\prime \tilde{\phi}_{n-1,C}+u^\prime\tilde{\phi}_{n,B}=\tilde{\phi}_{n,A} E(k), \label{normalized1} \\
&u^\prime\tilde{\phi}_{n,A}+v^\prime \tilde{\phi}_{n,C}=\tilde{\phi}_{n,B}E(k), \label{normalized2}\\ 
&v^\prime \tilde{\phi}_{n,B}+w^\prime\tilde{\phi}_{n+1,A}=\tilde{\phi}_{n,C} E(k), \label{normalized3}
\end{align}
where $\tilde{\phi}_{n,j}$ are renormalized flux velocity (see \ref{appendixa}  for details) and the new couplings are found as
\begin{align}
&u'=\sqrt{\frac{S_B S_C}{\left(S_{C}+S_{A}\right)\left(S_{A}+S_{B}\right)}},\label{closed couplings11}\\
&v'=\sqrt{\frac{S_A S_C}{\left(S_{B}+S_{A}\right)\left(S_{B}+S_{A}\right)}},\label{closed couplings22}\\
&w' =\sqrt{\frac{S_A S_B}{\left(S_{C}+S_{A}\right)\left(S_{C}+S_{B}\right)}}.\label{closed couplings33} 
\end{align}
By using Bloch's theorem once again, $\tilde{\phi}_{n,A} = \tilde{\phi}_{A}e^{iqnd}$, $\tilde{\phi}_{n,B} = \tilde{\phi}_{B}e^{iqnd}$, $\tilde{\phi}_{n,C}= \tilde{\phi}_{C}e^{iqnd}$,
we obtain the following eigenvalue problem for the flux velocity,

\begin{align}\label{acoustic_ssh3_flux}
   \begin{pmatrix}
   0 & u' & w'e^{-iqd}\\
   u' & 0 & v' \\
   w'e^{iqd} & v' & 0 
   \end{pmatrix} \begin{pmatrix}
   \tilde{\phi}_{A}\\
   \tilde{\phi}_{B}\\
   \tilde{\phi}_{C}
   \end{pmatrix} = E(k) \begin{pmatrix}
   \tilde{\phi}_{A}\\
   \tilde{\phi}_{B}\\
   \tilde{\phi}_{C}
   \end{pmatrix}.
\end{align}
One can observe that the mapping of the flux velocities is equivalent to changing the cross sections of the pressure mapping in the following manner: $S_{1}\rightarrow 1/S_{1}, S_{2}\rightarrow 1/S_{2}, S_{3}\rightarrow 1/S_{3}$ and BBC  will change accordingly as we will see in the next section. Furthermore, this formalism can be identified as a continuum version of strain coordinates \cite{allein2022strain}, recently introduced in topological mechanical metamaterials to describe the discrete version of zero velocity boundary conditions (free boundaries for a mechanical system). 
Thus, as long as the monomodal assumption holds, with either open or closed boundary conditions, the acoustic analog is an ideal ground to test experimentally predictions derived by BBC for the case of the SSH3. In Fig.\,\ref{fig2}\,\textbf{(c)}, we present the dispersion for the acoustic model obtained via Transfer Matrix Method (TMM) - circles - of the 1D monomodal approximation and the 2D problem solved with finite element methods (FEM) - red line. Notice that due to the "pseudo-energy" relation Eq. \ref{pseudoEnergy}, $kL=\pm\arccos{(E)}+2m\pi$, with positive $kL$, the $3$ propagating bands of the SSH3 model for $E$ (see Fig.\ref{fig1}\,\textbf{a})) are translated into an infinity of branches for $kL$. In addition, we see that, up to $8$\,kHz, the monomodal assumption is valid with very good precision (for more details see Appendix \ref{appendixb}).

 \par 

\subsection{Finite system and phase diagram}

\subsubsection{Bulk-boundary correspondence/ Phase diagram}
\begin{figure*}[ht] 
\begin{center}
\includegraphics[width=\textwidth]{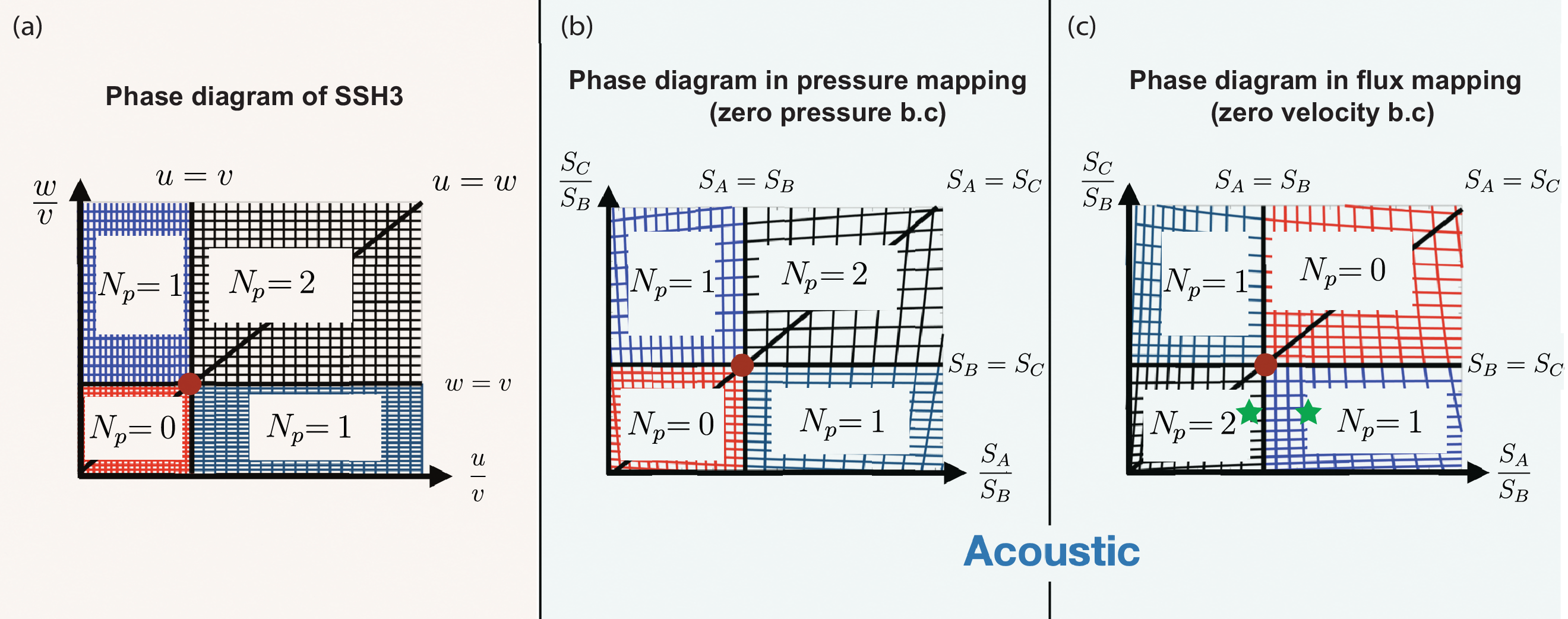}
 \caption{\label{fig4}\textbf{Phase diagram of SSH3} \textbf{a)} Phase diagram of an SSH3 according to normalized sublattice Zak phase. One can observe regions with 0,1 and 2 edge states. The bold black lines $u=w$, $w=v$ and $u=v$ represent mirror symmetric realizations of the system. We also present a grid with lines were either $u=\mathrm{const.}$ or $w=\mathrm{const.}$ and different colours for each region. We do this to highlight in \textbf{b)} and \textbf{c)} how this grid will be "distorted" under the mappings to the acoustic systems. \textbf{b)} The phase diagram of the acoustic analog in pressure mapping and zero pressure b.cs. One can observe that the regions are inverted with respect to the red disk (fully degenerate case - monomer). Furthermore the grid now acquires some curvature but interestingly the "mirror symmetric lines" remain intact under the mapping. \textbf{c)}. The phase diagram of the acoustic analog in flux mapping and zero velocity b.cs. The regions are inverted with respect to the pressure mapping and it resembles the phase diagram of the original SSH3. }
\end{center}
\end{figure*}


With the aim of further investigating the SSH3 mapping to the acoustic periodic waveguide, we now consider waveguides of finite size. In order to establish BBC for the SSH3, the amplitude of the field at the the ends of the finite structure should be zero. Thus, for a waveguide with ends opened to free space corresponding to zero pressure boundary conditions (b.cs), we use the pressure mapping. Similarly, in the case of an acoustic waveguide with ends closed by hard walls, corresponding to zero velocity b.cs, we use the flux velocity mapping.


In Fig.\,\ref{fig4} we demonstrate how the phase diagram of SSH3 is mapped in each case (acoustic system with zero velocity or zero pressure b.cs). In  particular, illustrated in Fig.\,\ref{fig4}\textbf{(a)} is the phase diagram for an SSH3 with $3N+1$ sites as predicted by $Z_{\lambda_i}^{B}$. Four different parametric regions can be identified: one region with no edge states (when $v >u,w$), two regions with one pair of edge states ($w>v >u$ or $u>v >w$) and one with two pairs of edge states ($v< u,w$). For the acoustic system (Fig.\,\ref{fig4}\,\textbf{(b-c)}), the relations between the couplings is translated to relations between the cross-sections as indicated by \eqref{open couplings1}-\eqref{open couplings3} and \eqref{closed couplings11}-\eqref{closed couplings33}. Interestingly, the phases are interchanged between the different mappings. One can observe that if one system with open ends (zero pressure b.cs) supports two pairs of edge states, then these edge states will "disappear" if the same system is closed (zero flux velocity b.cs).
 On the contrary we expect to see two pairs of edge states for $S_A, S_C > S_B$ for the closed system. 
We proceed with the numerical demonstration of our results for two waveguide setups, one with open ends and one with closed ends.

\subsubsection{Numerics}

\begin{figure*}[ht] 
\begin{center}
\includegraphics[width=0.85\textwidth]{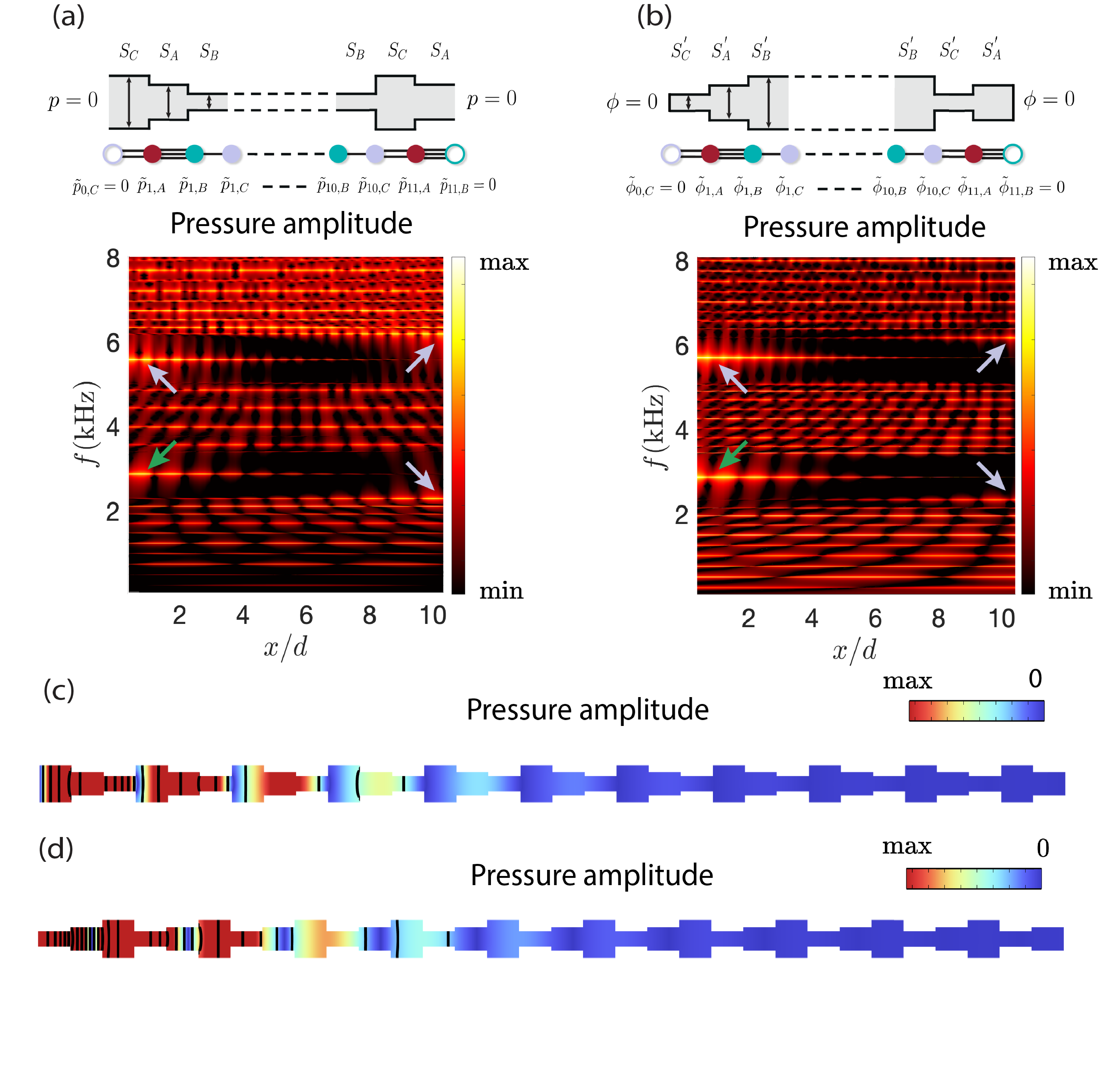}
 \caption{ \textbf{Numerical result of zero pressure vs zero velocity b.cs for different bulks}.  \textbf{(a)}  (top) Schematic for the SSH3 lattice with zero pressure b.cs. The three cross-sections are chosen $S_A=0.75$\,cm$^2$, $S_B=0.5$\,cm$^2$ $S_C=1.2$\,cm$^2$ corresponding to couplings $u=0.4804$, $v=0.3430$  and $w=0.6591$. (bottom) Spatial profiles of the pressure field corresponding to each frequency  using FEM with zero pressure b.cs. Two edge states are exhibited in each band-gap, indicated by arrows. 
 \textbf{(b)}  (top) Schematic for the SSH3 lattice with zero flux velocity b.cs. The three cross-sections are chosen $S_A^\prime=0.75$\,cm$^2$, $S_B^\prime=1.2$\,cm$^2$ $S_C^\prime=0.5$\,cm$^2$ corresponding to couplings $u^\prime=0.4961$, $v^\prime=0.3363$  and $w^\prime=0.6508$. (bottom)
 Spatial profiles of the pressure field corresponding to each frequency using FEM with zero pressure b.cs. Two edge states are exhibited in each band-gap, indicated by arrows. 
 In \textbf{c)} and \textbf{d)}, we present the 2D profile of the pressure of the edge states indicated by the green arrows in \textbf{(a)} and \textbf{(b)}. The lines correspond to the contour of the pressure wavefront. \label{fig3}}
\end{center}
\end{figure*}

To verify the theoretical predictions of Fig.\,\ref{fig4} we solve the 2D lossless Helmholtz equation using finite element methods (FEM) for a finite size system. Following \cite{coutant2021acoustic}, we consider open (zero pressure b.cs) and closed systems (zero flux velocity b.cs) as illustrated in Fig.\,\ref{fig3}\,\textbf{(a)} and \textbf{(b)}. 

The length of each waveguide segment is chosen $L=2$\,cm, while the three cross-sections are $S_A=0.75$\,cm$^2$, $S_B=0.5$\,cm$^2$, $S_C=1.2$\,cm$^2$ corresponding to couplings $u=0.4804$, $v=0.3430$ and $w=0.6591$ \eqref{open couplings1}-\eqref{open couplings3} for the waveguide configuration with zero pressure b.cs depicted in Fig.\,\ref{fig3}\,\textbf{(a)}. The number of sites is 31 (see Fig.\,\ref{fig3}\,\textbf{(a)}), which in return fixes the total length of the waveguide to $l=64$\,cm. For this system, the SSH3 model predicts two edge modes for each gap. 
The corresponding simulated pressure field for each site is presented on the bottom of Fig.\,\ref{fig3}\,\textbf{(a)}. One can clearly observe the different mode profiles of the pressure at discrete frequencies of the spectrum and the formation of three propagating bands. In addition to these propagating bands, on can observe two distinct localized modes in each bandgap as predicted by the SSH3 phase diagram in Fig.\,\ref{fig4}\textbf{(b)}. The frequencies of two first edge modes from 2D FEM simulation are respectively $f^{(n)}_{1,op} = 2.298$\,kHz and $f^{(n)}_{2,op}=2.876$\,kHz, very close to the analytical predictions given by $f^{(a)}_{1,op}\approx c_{0}/(2\pi L)\arccos{(w)}=2.323$\,kHz and $f^{(a)}_{2,op}\approx c_{0}/(2\pi L)\arccos{(u)}=2.919$\,kHz, assuming that the speed of sound is $c_0=343$\,m/s. 
In Fig.\,\ref{fig3}\,\textbf{(c)} we demonstrate the localization of the second edge mode with frequency $f^{(n)}_{2,op}$ by showing the 2D pressure profile, where the lines correspond to the contour of the pressure field. Focusing in the region close to the cross-section change, one can observe some weak 2D effects, indicating only a slight deviation from a completely planar waveform. Despite this, the 2D FEM simulations validate the 1D discrete approximation of the acoustic system.

Next, we change from zero pressure to zero flux velocity b.cs on both ends as depicted in Fig.\,\ref{fig3}\,\textbf{(b)} and we choose a different structure with  $S_A^\prime=0.75$\,cm$^2$, $S_B^\prime=1.2$\,cm$^2$ $S_C^\prime=0.5$\,cm$^2$ corresponding to the prime couplings $u^\prime=0.4961$, $v^\prime=0.3363$ and $w^\prime=0.6508$ for which the SSH3 model predicts two edge modes for each gap. 
Once again the numerical simulation shows an emergence of three distinct propagating bands where the energy is concentrated and the existence of two edges modes per band gap as illustrated by arrows (see Fig.\,\ref{fig3}\,\textbf{(b)}).
One can notice that although the structures differ, they are both in the two edge state topological phase as predicted by Fig.\ref{fig4}\,\textbf{(b)} and \textbf{(c)}. The frequencies of two first edge modes from 2D FEM simulation are respectively $f^{(n)}_{1,cl} = 2.309$\,kHz and $f^{(n)}_{2,cl}= 2.837$\,kHz, again very close to the analytical ones given by $f^{(a)}_{1,cl}\approx c_{0}/(2\pi L)\arccos{(w')}=2.353$\,kHz and $f^{(a)}_{2,cl}\approx c_{0}/(2\pi L)\arccos{(u')}=2.870$\,kHz. Finally, in Fig.\,\ref{fig3}\,\textbf{(d)} we present the 2D pressure profile of the second edge mode, as in the Fig.\,\ref{fig3}\,\textbf{(c)}. 

As anticipated, the numerical simulation validates the theoretical predictions of the phase diagram in Fig.\ref{fig4} for the corresponding values of the cross sections for both zero flux velocity and zero pressure b.cs.

\section{Experimental results}

\begin{figure*}[ht]
\begin{center}
\includegraphics[width=1\textwidth]{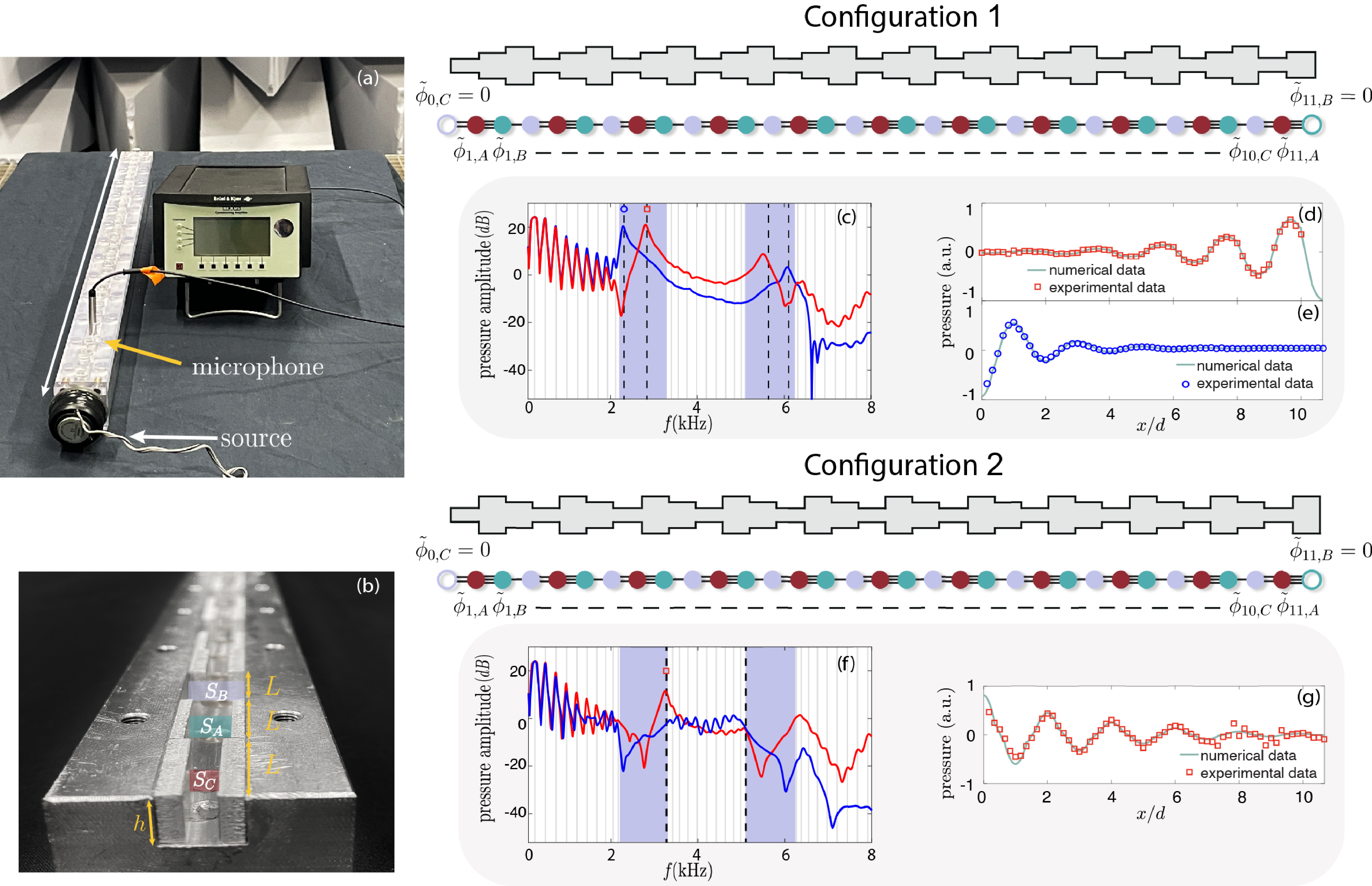}
 \caption{\textbf{Experimental realization} \textbf{(a)} Experimental setup. A side view of the total waveguide of length $l=64$\,cm composed of 31 changes of cross-section, closed with the plexiglass plate on top. The microphone and the source are mounted into the system.
\textbf{(b)} A closeup view of the waveguide. The three different cross-sections $S_{A,B,C}$, the periodicity $3L$ and the height $h$ of the rectangular waveguide are indicated. \textbf{(c)} Pressure amplitude as a function of frequency for the SSH3,  for the waveguide configuration illustrated on the top (configuration 1). Below the waveguide we present its discrete flux representation. The couplings in this case are  $u^\prime=0.4961$ , $v^\prime=0.3363$ and $w^\prime= 0.6508$, and the waveguide maps to a SSH3 configuration with 2 edge modes per bandgap. The dashed black (grey) vertical lines correspond to the frequency of the edge (non-localized) modes as obtained numerically. \textbf{(d)} The pressure profile of the edge mode localized on the left hand side of the setup, from numerics in ciel (line) and from experiments with red square. The frequency of the experimental profile is indicated  by a red square in panel \textbf{(c)}. \textbf{(e)} Same as in \textbf{(d)} for the second edge mode localized on the right boundary. The frequency of the experimental profile is indicated by a blue circle in panel \textbf{(c)}. 
 \textbf{(f)} Same as panel \textbf{(c)} but with a different waveguide configuration as illustrated on the  bottom (configuration 2). The couplings in this case are given as $u^\prime=0.3363$, $v^\prime=0.4961$  and $w^\prime=0.6508$, which maps to a SSH3 configuration with a single edge mode per bandgap. The dashed black (grey) vertical lines correspond to the frequency of the edge (non-localized) modes as obtained numerically.
\textbf{(g)}  Similar as in \textbf{(d)} and \textbf{(e)} for the edge mode of the first bandgap. The frequency of the experimental profile is indicated by a red square in panel \textbf{(f)}.
 \label{fig5}}
\end{center}
\end{figure*}

The experimental setup is based on an acoustic rectangular waveguide with periodically alternating change of cross-section,   as depicted in Fig.\,\ref{fig5}\,\textbf{(a)} and \textbf{(b)}. The length of the waveguide is $l=64$\,cm and the changes of cross-section occur every $L=2$\,cm which fixes the changes of cross-section to 31. The periodic arrangement of the segments is achieved by adding aluminium blocks similarly to \cite{coutant2021acoustic}. The waveguide is closed by screwing a plexiglass plate on top of it. The acoustic pressure is measured with microphones at different positions on the bottom of the waveguide and on the plexiglass top plate as depicted in Fig.\,\ref{fig5}\,(a).

As mentioned in the previous section, the leakage due to the open ends of the experimental setup  induces additional dissipation, thus we choose to work with closed boundaries (zero velocity boundary conditions). This is realized by skrewing two plexiglass plates at the two ends of the waveguide. 
We perform experiments with two different waveguide configurations, one corresponding to the region of the phase diagram with 2 edge modes (black grid) and one for the region with 1 edge mode (blue grid) at each gap respectively (see Fig.\,\ref{fig4}\,\textbf{(c))}.
The signal source used in the experiments is a sweep-sine adjusted from 100\,Hz to 8\,kHz. As mentioned in the numerical simulations, when two edge modes exist in the same bandgap, they are localized at the opposite ends of the waveguide. Hence,
two different experiments are conducted, each with a source positioned at opposite end.

Depicted with red line in Fig.\,\ref{fig5}\,\textbf{(c)} is the spectrum measured at 2\,cm from the source located at the left of the structure for the configuration illustrated on the top (configuration 1). We have chosen the same cross-sections as in the numerical simulations of the previous section, which we repeat  here for convenience, $S_A=0.75\times1$\,cm$^2$, $S_B=1.2\times1$\,cm$^2$,  $S_C=0.5\times1$\,cm$^2$.
This configuration maps to a SSH3 lattice with 2 edge modes in each gap with couplings corresponding to $u^\prime=0.4961$, $v^\prime=0.3363$ and $w^\prime=0.6508$. The blue line corresponds to the spectrum when the source is positioned on the right end of the waveguide and the microphone at 2 cm away. In pale violet color, we illustrate the bandgap of the infinite periodic system. The two edge modes in each gap (four in total) are clearly visible as indicated by the large resonance peaks inside the bandgaps. The black dashed vertical lines correspond to the frequency of each edge mode as predicted by FEM simulation of the finite structure. 
We remind that the analytical predictions for the frequencies of the edge modes in the first bandgap are found to be  $f^{(a)}_{1,cl}\approx 2.353$\,kHz and $f^{(a)}_{2,cl}\approx 2.870$\,kHz, while the numerical values are found to be $f^{(n)}_{1,cl}=2.309$\,kHz and $f^{(n)}_{2,cl}=2.837$\,kHz respectively. Finally, the experimentally obtained frequencies are found $f^{(e)}_{1,cl}=2.270$\,kHz and $f^{(e)}_{2,cl}=2.813$\,kHz indicating that the experimental/numerical results and the analytical predictions are in very good agreement.

To further solidify the predictions of the SSH3 model, we also present the experimental pressure profiles at the frequency of the peaks inside the first bandgap for the two source configurations. In Fig.\,\ref{fig5}\,\textbf{(d)} and \textbf{(e)}, we show the profiles corresponding to the frequency indicated by a red square or blue circle in panel \textbf{(c)} respectively. The blue line corresponds to the edge modes found numerically at the frequencies corresponding to the dashed black lines of the panel (c).
The experimental profiles are in excellent agreement with the numerical predictions. Moreover, it is clearly shown that these two modes in the bandgap correspond to edge modes of the system corroborating the theoretical prediction.

Fig.\,\ref{fig5}\,\textbf{(f)} shows the frequency spectrum measured at 2 cm from the source located at the left of the structure illustrated by configuration 2. The cross-sections are $S_A=1.2\times1$\,cm$^2$, $S_B=0.75\times1$\,cm$^2$,  $S_C=0.5\times1$\,cm$^2$ corresponding to couplings $u^\prime=0.3363$, $v^\prime=0.4961$  $w^\prime=0.6508$, which map to a SSH3 with a single edge state per bandgap. The edge mode in the first bandgap is located at 3.25 kHz, in the vicinity to the higher cutoff frequency of the first bandgap. The analytically predicted eigenfrequency of the edge state in the first bandgap is $3.351$\,kHz in very good agreement with the experimental value. 
Finally, in Fig.\,\ref{fig5}\textbf{(g)}, we present also the experimental pressure profile of the edge mode corresponding to the frequency indicated by a red triangle in panel \textbf{(f)} and we compare it with pressure profile obtained by the FEM simulation.\\


\section{Conclusions}

In summary, we have established an acoustic analogue of the SSH3 model without chiral and mirror symmetry, for both acoustic pressure and flux velocity fields. In particular, for finite systems, we have shown that the choice of the dependent variable is essential to predict the number of edge states of the system, relying on the boundary conditions. By performing experiments using acoustic waveguides, we explored the different regimes of the edge state phase diagram. Our analysis extends the findings of \cite{allein2022strain} beyond mass-spring models to a class of continuous systems consisting of periodic arrangement of slender waveguides. It is a challenging question if such an analysis can be applied to a wider class of continuous systems and with more complex, tailored boundaries.



 


\bibliography{bibliography}

\appendix

\section{Edge modes of the $3N+2$ finite SSH3 lattice}\label{appendix0}

\begin{figure*}[ht]
\begin{center}
\includegraphics[width=0.9\textwidth]{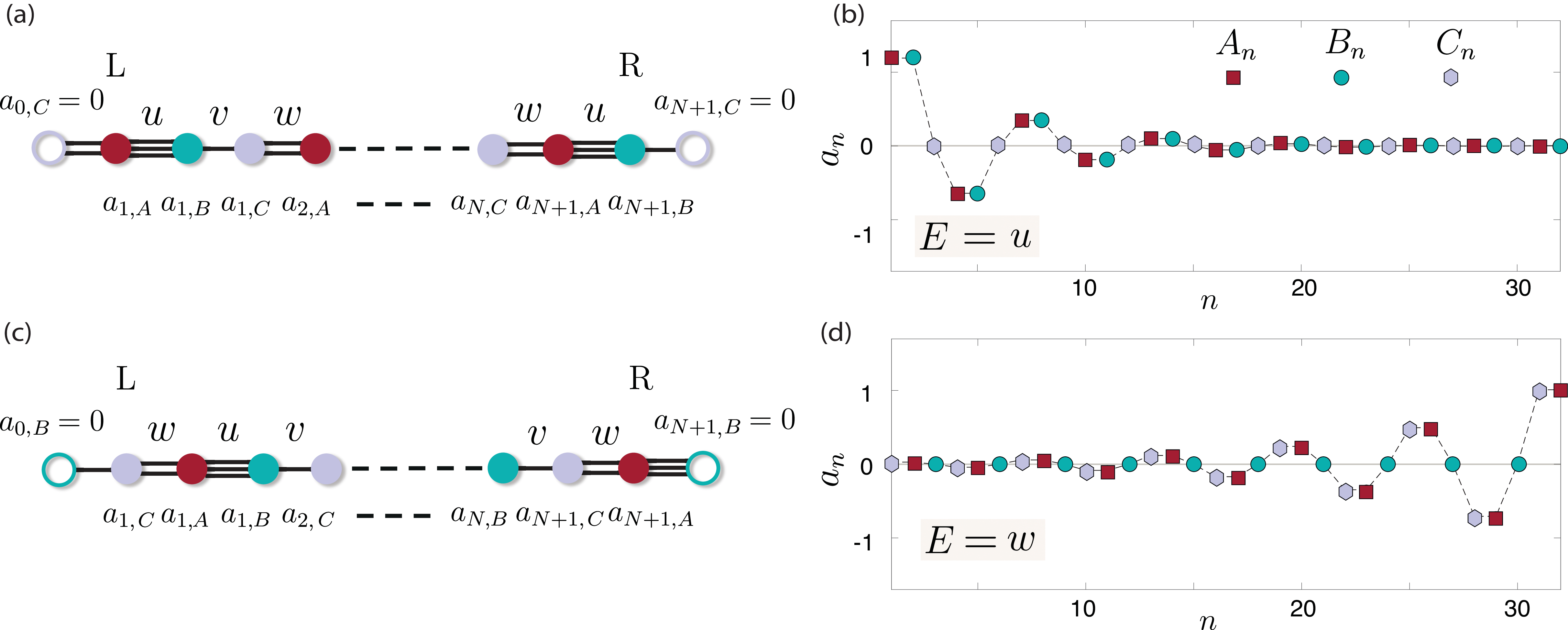}
 \caption{\textbf{Finite SSH3 lattice with 3$N$+2 sites.} A finite (open) SSH3 lattice with 3$N$+2
sites and two different terminations and their corresponding edge modes for $u=0.4804,\,w=0.6591$ and $v=0.3430$. 
 \label{fig6}}
\end{center}
\end{figure*}
\begin{figure}[ht]
\begin{center}
\includegraphics[width=0.4\textwidth]{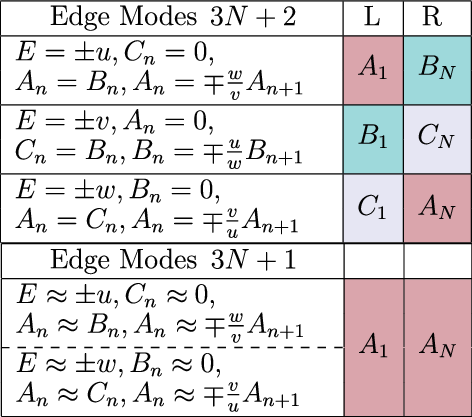}
 \caption{\textbf{Characteristics of the edge modes of the SSH3 chain.} On the top, the three different scenarios depending on the terminations of the $3N+2$ chain. On the bottom, the edge mode characteristics of the $3N+1$ chain, corresponding to the scenario chosen for the numerical and experimental demonstration.
 \label{fig7}}
\end{center}
\end{figure}

The system of equations for each lattice site $a_{n,A}=A_n$, $a_{n,B}=B_n$ and $a_{n,C}=C_n$, for a finite SSH3 chain is
\begin{align}
&u B_{n}+w C_{n-1}=E_n A_{n},\label{infinite1}\\ 
&v C_n+u A_n=E_n B_n, \label{infinite2}\\
&v B_n+w A_{n+1} =E_n C_n.\label{infinite3}
\end{align}
In order to determine the edge mode characteristics (energy, localization and profile) of the SSH3, we impose vanishing boundary conditions on the field at the edges of the lattice as depicted in Fig.\,\ref{fig6}. 
For a lattice composed of $3N+2$ sites, after some straightforward algebraic manipulations, we find exact analytical solutions for the edge modes. The results for the three different configurations, depending on which sub-lattice the first site belongs, are summed up in Fig\,\ref{fig7}.



Concurrently, each region near the edges of the $3N$ and $3N+1$ chains may be viewed as a semi-infinite $3N+2$ chain, as long as $N$ is not small. It is now apparent that the edge modes of the $3N$ and $3N+1$ may be approximated by the edge mode solutions of the $3N+2$ chains. For the $3N+1$ termination chosen in this article (both left and right edges starting with the sublattice $A_n$), we present the approximate edge mode characteristics on the bottom of Fig\,\ref{fig7}.

\section{Discrete approximation of the waveguide system} \label{appendixa}

\subsection{Pressure mapping}
In this section, following the same line as in \cite{coutant2021acoustic}, we showcase how one can obtain equations \eqref{normalized1} to \eqref{normalized3} by solving the 1D Helmholtz equation and applying the proper boundary conditions for each waveguide segment. Specifically, at each change of cross-section, the jump conditions \eqref{jump_cond_1} should be satisfied. By solving the 1D Helmholtz equation for two waveguide segments (one between $A$ and $B$ and one between $B$ and $C$, denoted as $ABC$), one gets:
\begin{align}
    p_{n,A} & = p_{n,B}\cos{kL} - \frac{p^\prime_{n,B^-}}{k}\sin{kL}\\
    p_{n,C} & = p_{n,B}\cos{kL} + \frac{p^\prime_{n,B^+}}{k}\sin{kL},
\end{align}
where $p^\prime_{n,j} := S_j \frac{d p_{n,j}}{dx}$. By multiplying the first equation by $S_{A}$, the second by $S_{B}$ and adding them up while implementing the jump condition, one obtains:
\begin{align}
    \frac{S_{A}}{S_{A}+S_{B}}p_{n,A} + \frac{S_{B}}{S_{A}+S_{B}}p_{n,C} = p_{n,B}\cos{(kL)}
\end{align}
In a similar way, by solving the Helmholtz equation and applying the boundary conditions for the segments $BCA$ and $CAB$, one gets the following system of equations:
\begin{align}
&\frac{S_{C}}{S_{C}+S_{A}} p_{n-1,C}+\frac{S_{A}}{S_{C}+S_{A}} p_{n,B}=p_{n,A} \cos{(kL)}, \label{discrete1}\\
&\frac{S_{A}}{S_{A}+S_{B}}p_{n,A}+\frac{S_{B}}{S_{A}+S_B} p_{n,C}=p_{n,B}\cos{(kL)},
 \label{discrete2} \\
&\frac{S_{B}}{S_{B}+S_{C}} p_{n,B}+\frac{S_{C}}{S_{B}+S_{C}} p_{n+1,A}=p_{n,C} \cos{(kL)}.\label{discrete3}
\end{align}
At first glimpse the resulting Hamiltonian will not be symmetric. However we can normalize pressure following the normalisation proposed in \cite{coutant2021acoustic} i.e
\begin{align}
\tilde{p}_{n,A} &=\sqrt{S_{C}+S_{A}} p_{n,A}  \\
\tilde{p}_{n,B}&=\sqrt{S_{A}+S_{B}} p_{n,B},\\
\tilde{p}_{n,C} &=\sqrt{S_{B}+S_{C}} p_{n,C},
\end{align}
which we substitute in Eqs\,(\ref{discrete1}-\ref{discrete3}) and obtain equations \eqref{normalized1} - \eqref{normalized3}, for which the Hamiltonian is Hermitian.

\subsection{Flux mapping}
In contrast to the pressure mapping established in the previous section, we now show that an equivalent mapping may be found for the flux velocity field. Starting from Eq.\,\eqref{Helmholtz_1d}, 
the transfer matrix for each segment with a constant cross section is:

\begin{equation}\label{transf-mat}
T_l=
\left[\begin{array}{ll}
\cos \left(kL\right)&\mathrm{i}Z_l\sin\left( kL\right)\\
\frac{\mathrm{i}}{Z_l}\sin \left(kL\right)&\,\,\,\,\,\cos \left(kL \right)
\end{array}\right],
\end{equation}
where $Z_l=\rho_0 c_0/ S_l$ is the waveguide impedance, $\rho_0$ is the density of air and the subscript $l=CA,AB,BC$ denotes the corresponding waveguide segments. Taking the left and right propagation of flux at each change of cross section and applying the jump conditions \eqref{jump_cond_1}, we get the following relations for the pressure-flux amplitudes at for the segments $AB$ and $BC$:

\begin{align}
    &\phi_{n,B}= \phi_{n,A}\cos{(kL)}+\frac{\mathrm{i}}{Z_A}\sin{(kL)}p_{n,A},\label{system1}\\
    &\phi_{n-1,C}= \phi_{n,A}\cos{(kL)}-\frac{\mathrm{i}}{Z_C}\sin{(kL)}p_{n,A}.\label{system2}
\end{align}
Next we multiply \eqref{system1} by $Z_A$ and \eqref{system2} by $Z_C$ and add them up. By dividing the resulting equation with $1/(Z_{C}+Z_{A})$, we get:

\begin{equation}
  \frac{Z_{C}}{Z_{C}+Z_{A}} \phi_{n-1,C}+\frac{Z_{A}}{Z_{C}+Z_{A}} \phi_{n,B}=\phi_{n,A} \cos{(kL)}. \label{discrete closed1}
\end{equation}
By repeating the same procedure with the rest of the segments, we arrive at the following system of equations that describe a discrete SSH3 model: 

\begin{align}
&w \tilde{\phi}_{n-1,C}+u \tilde{\phi}_{n,B}=\tilde{\phi}_{n,A} E(k), \label{normalizedap1} \\
&u\tilde{\phi}_{n,A}+v \tilde{\phi}_{n,C}=\tilde{\phi}_{n,B}E(k), \label{normalizedap2}\\ 
&v \tilde{\phi}_{n,B}+w\tilde{\phi}_{n+1,A}=\tilde{\phi}_{n,C} E(k), \label{normalizedap3} \\ \nonumber
\end{align}
where: 
\begin{align}
&\tilde{\phi}_{n,A} =\sqrt{Z_{C}+Z_{A}} \phi_{n,A}, \\
&\tilde{\phi}_{n,B}=\sqrt{Z_{A}+Z_{B}} \phi_{n,B}, \\& \tilde{\phi}_{n,C} =\sqrt{Z_{B}+Z_{C}} \phi_{n,C},
\end{align}
and:
%

\begin{align}
&u'=\frac{Z_A}{\sqrt{\left(Z_{C}+Z_{A}\right)\left(Z_{A}+Z_{B}\right)}}=\sqrt{\frac{S_B S_C}{\left(S_{C}+S_{A}\right)\left(S_{A}+S_{B}\right)}}\label{closed couplings111},\\
&v'=\frac{Z_B}{\sqrt{\left(Z_{B}+S_{C}\right)\left(Z_{A}+Z_{B}\right)}}=\sqrt{\frac{S_A S_C}{\left(S_{B}+S_{A}\right)\left(S_{B}+S_{A}\right)}},\label{closed couplings222}\\
&w' =\frac{Z_C}{\sqrt{\left(Z_{C}+Z_{A}\right)\left(Z_{C}+Z_{B}\right)}}=\sqrt{\frac{S_A S_B}{\left(S_{C}+S_{A}\right)\left(S_{C}+S_{B}\right)}}\label{closed couplings333}.
\end{align}
%



\section{Dispersion relation of the periodic waveguide} \label{appendixb}

Since we have limited our analysis in the lossless and low frequency regime, such that only the plane acoustic mode is propagating, one can use the transfer matrix and the Bloch theorem for a periodic unit cell of the trimer waveguide, as the one depicted in Fig.\ref{fig2}\textbf{a)}, in order to obtain the dispersion relation of the infinite periodic system.  The pressure and the flux velocity between the sites $A_{n+1}$ and $C_{n}$, between the sites $C_{n}$ and $B_{n}$ and between the sites $B_{n}$ and $A_{n}$ and inside the waveguide, each separated by a distance $L$  are connected through the transfer matrix \eqref{transf-mat}. Thus, the pressure and the flux velocity between the sites $A_n$ and $A_{n+1}$ inside the waveguide, separated by a distance $d=3L$, are connected 
with the following transfer matrix
\begin{align}
\left[\begin{array}{ll}
 p_{n+1,A}\\
 \phi_{n+1,A}
\end{array}\right]=T\left[\begin{array}{ll}
  p_{n,A}\\
 \phi_{n,A}
\end{array}\right]=T_C T_BT_A\left[\begin{array}{ll}
  p_{n,A}\\
 \phi_{n,A}
\end{array}\right], \label{transfer}
\end{align}

Assuming an infinite periodic waveguide with translation invariance, Eq. (\eqref{transfer}) must hold for any $n$. Thus we may solve this equation by using a Bloch like anstaz \cite{soukoulis}. According to this ansatz the state vector will be given by
\begin{equation}
    \left[\begin{array}{ll}
 p_{n,A}\\
 \phi_{n,A}
\end{array}\right]=A\mathrm{e}^{\mathrm{i}qnd },
\end{equation}
 where $A$ is a column eigenvector.
Substituting this expression into equation \eqref{transfer} we obtain the following eigenvalue equation for the unknown phase factor $qd$
 \begin{equation}  TA=\mathrm{e}^{\mathrm{i}qd}IA. \label{eigevalue eq}
\end{equation}
Hence, $\mathrm{e}^{\mathrm{i}qd}$ is an eigenvalue of the transfer matrix $T$. Due to time reversal symmetry, the determinant of the transfer matrix is equal to unity. As a result its eigenvalues $\lambda_{1,2}$ satisfy the following expression
\begin{equation}
   \det{(T)}=\lambda_1 \lambda_2 =1 \label{det}.
\end{equation}
Then, one can identify that
 \begin{equation}
   \Tr(T)=2\cos{(qd)}. \label{trace}
\end{equation}
From the latter expression one can obtain the dispersion relation of the periodic acoustic waveguide, i.e

\begin{align}
    &\cos (q d)=\cos ^{3}(k L)-Q\cos (k L)\sin ^{2}(k L), 
    \label{transfer matrix}
\end{align}
where   
\begin{align}
Q=\frac{{Z_A}^2\left(Z_B+Z_C\right)+Z_A\left({Z_B}^2+{Z_C}^2\right)+Z_BZ_C\left(Z_B+Z_C\right)}{{2Z_AZ_BZ_C}} \nonumber
\end{align}

In order to verify the 1D approximation, we compute the dispersion of the periodic waveguide by solving the lossless 2D Helmholtz equation (\ref{helmholtz}) using  finite element methods (FEM) and we compare it with expression (\ref{transfer matrix}) derived through the TMM. The results are shown in Fig.\,\ref{fig2}\textbf{(c)}, where the red lines (solid) correspond to the FEM  while the black circles correspond to the TMM. The FEM simulations are in excellent agreement with Eq.\,(\ref{transfer matrix}) until the three first propagating bands and above this region is less accurate.

\end{document}